\documentclass[preprint,aps]{revtex4}

\usepackage{graphicx}
\usepackage{dcolumn}
\usepackage{bm}

\def\x{{\bf x}}

\def\k{\mathbf{k}}
\def\q{\mathbf{q}}
\def\a{\alpha}
\def\b{\beta}
\def\la{\langle}
\def\ra{\rangle}
\def\ria{\rightarrow}
\def\a{\alpha}
\def\au{{\underline{\alpha}}}
\def\s{{\sigma}}
\def\half{\frac{1}{2}}

\begin{document}


\title{Decoherence of Histories and Hydrodynamic Equations \\ for a Linear Oscillator Chain}

\author{J.J.Halliwell}%
\affiliation{Blackett Laboratory \\ Imperial College \\ London SW7
2BZ \\ UK }



\begin{abstract}
We investigate the decoherence of histories of local densities for
linear oscillators models. It is shown that histories of local
number, momentum and energy density are approximately decoherent,
when coarse-grained over sufficiently large volumes. Decoherence
arises directly from the proximity of these variables to exactly
conserved quantities (which are exactly decoherent), and not from
environmentally-induced decoherence. We discuss the approach to
local equilibrium and the subsequent emergence of hydrodynamic
equations for the local densities.

\vskip 1.0in
\centerline{Imperial/TP/2-03/23, May 2003}

\end{abstract}

\maketitle

\section{Introduction}

In a large and possibly complex quantum system, which dynamical
variables naturally become classical for a wide variety of initial
states? This question belongs to the general issue of emergent
classicality from quantum theory, and has recently received a
considerable amount of attention (see Ref.\cite{Har6} for an overview).
There are a number of different approaches to it, but common to
most of them is the demonstration of decoherence: that certain
types of quantum states of the system in question exhibit
negligible interference, and therefore superpositions of them are
effectively equivalent to statistical mixtures.

Decoherence has principally been demonstrated for the situation in
which there is a distinguished system, such as a particle, coupled
to its surrounding environment \cite{JoZ,Zur}. More generally, we
may expect that decoherence comes about when the variables
describing the entire system of interest naturally separate into
``slow'' and ``fast'', whether or not this separation corresponds
to, respectively, system and environment. (See Ref.\cite{BrHa} for
a discussion of the conditions under which the total Hilbert space
may be  written as a tensor product of system and environment
Hilbert spaces). If the system consists of a large collection of
interacting identical particles, as in a fluid for example, the
natural set of slow variables are the local densities: energy,
momentum, number, charge {\it etc.} These variables, in fact,
are also the variables which provide the most complete description
of the classical state of a fluid at a macroscopic level. The most
general demonstration of emergent classicality therefore consists
of showing that, for a large collection of interacting particles
described microscopically by quantum theory, the local densities
become effectively classical. Although decoherence through the
system--environment mechanism is expected to play a role since
the collection of particles are coupled to each other, it is of
interest to explore the possibility that, at least in some regimes,
decoherence could come
about for a different reason. Namely, because
the local densities are almost conserved if averaged over a
sufficiently large volume \cite{GH2}. Hence, the approximate
decoherence of local densities would then be due to the fact
that they are close to a set of exactly conserved quantities, and
exactly conserved quantities obey superselection rules.

We will approach the question using the decoherent histories
approach to quantum theory \cite{GH2,GH1,Gri,Omn,Hal1,Hal5}. This approach
has proved particularly useful for discussing emergent
classicality in a variety of contexts. In particular the
issues outlined above are most clearly expressed in the language
of decoherent histories. The central object of interest is the
decoherence functional,
\begin{equation}
D (\au, \au') = {\rm Tr} \left( P_{\a_n} (t_n) \cdots P_{\a_1}
(t_1) \rho P_{\a_1'} (t_1) \cdots
P_{\a_n'} (t_n) \right)
\label{1.1}
\end{equation}
The histories are characterized by the initial state $ | \Psi \ra
$ and by the strings of projection operators $P_{\a} (t)$
(in the Heisenberg picture) at times
$t_1$ to $t_n$ (and $\au$ denotes the string of alternatives $\a_1
\cdots \a_n$). Intuitively, the decoherence functional is a
measure of the interference between pairs of histories $\au$,
$\au'$. When it is zero for $\au \ne \au' $, we say that the
histories are decoherent and probabilities  $ p (\au ) = D (\au,
\au ) $ obeying the usual probability sum rules may be assigned to
them. One can then ask whether these probabilities are strongly
peaked about trajectories obeying classical equations of motion.
For the local densities, we expect that these equations will be
hydrodynamic equations.

We are generally concerned with a system of $N$ particles
interacting through a potential and are therefore described at the
microscopic level by a Hamiltonian of the form
\begin{equation}
H = \sum_j
{ {\bf p}_j^2 \over 2 m } + \sum_{\ell >j } V_{j\ell} (
\q_j - \q_\ell )
\end{equation}
We are particularly interested in the number density $n(\x)$, the
momentum density ${\bf g}(\x)$ and the energy density $ h (\x )$,
defined by,
\begin{eqnarray}
n(\x) &=& \sum_j \ \delta(\x -\q_j)
\\
{\bf g}(\x) &=& \sum_j \ {\bf p}_j \ \delta (\x- \q_j)
\\
h(\x) &=& \sum_j \  { {\bf p}^2_j \over 2 m } \delta(\x- \q_j) + \sum_{\ell >
j} V_{j \ell} (\q_j - \q_{\ell} ) \delta(\x- \q_j)
\end{eqnarray}
We are generally interested in the integrals of these quantities over small
volumes, which will have the effect of smearing out the $\delta$-functions.
Integrated over an infinite volume, these become the total particle
number $N$, total momentum $P$ and total energy $H$, which are exactly
conserved.
It is also often useful to work with the Fourier transforms of the
local densities,
\begin{eqnarray}
n(\k) &=& \sum_j \ e^{i \k \cdot \q_j}
\\
{\bf g}(\k) &=& \sum_j \ {\bf p}_j \ e^{i \k \cdot \q_j}
\\
h(\k) &=& \sum_j \  { {\bf p}^2_j \over 2 m} e^{i \k \cdot \q_j}+
\sum_{\ell > j} V_{j \ell} ( \q_j - \q_{\ell} ) \ e^{i \k \cdot
\q_j}
\end{eqnarray}
These quantities tend to the exactly conserved quantities in the
limit $ k = | \k | \ria  0 $.

There is a standard technique for deriving hydrodynamic equations
for the local densities \cite{Hua,hydro}. It starts with the continuity equations
expressing local conservation, which have the form,
\begin{equation}
\frac{ \partial \sigma} { \partial t} + \nabla \cdot {\bf  j } = 0
\label{1.9}
\end{equation}
where $\sigma $ denotes $n$, ${\bf g}$ or $h$ (and the current
${\bf j}$ is a second rank tensor in the case of ${\bf g}$).
It is then assumed that, for a wide variety of initial states,
conditions of local equilibrium are
established after a short period of time.
This means that on scales small compared to the overall
size of the fluid, but large compared to the microscopic scale,
equilibrium conditions are reached in each local region,
characterized by a local temperature, pressure {\it etc.} which vary
slowly in space and time. Local equilibrium is described by
the density operator
\begin{equation}
\rho = Z^{-1} \exp \left( - \int d^3x \ \b (\x) \left[ h(\x) - \bar \mu (\x)
n (\x) - {\bf u} (\x) \cdot {\bf g} (\x) \right] \right)
\label{1.10}
\end{equation}
where $ \b  $, $\bar \mu $ and $ {\bf u} $ are Lagrange multipliers and
are slowly varying functions of space and time. $\b$ is the inverse
temperature, ${\bf u}$ is the average velocity field, and $\bar \mu$
is related to the chemical potential which in turn is related to
the average number density. (Note that the local equilibrium state is defined in
relation to a particular coarse-graining, here, the anticipated calculation
of average values of the local densities. Hence it embraces all possible states
that are effectively equivalent to the state Eq.(\ref{1.10}) for the purposes of calculating
those averages.) The hydrodynamic
equations follow when the continuity equations are averaged in this
state. These equations form a closed set because the local equilibrium
form depends (in three dimensions) only on the five Lagrange multiplier
fields $ \b, \bar \mu $, $ {\bf u}$ and there are exactly five continuity equations
(\ref{1.9}) for them. (More generally, it is possible to have closure up to a set
of small terms which may be treated as a stochastic process. See Refs.\cite{BrHa,CaH1},
for example.)

The decoherent histories approach to quantum theory offers the possibility
of a much more general derivation of emergent classicality than that
entailed in the standard derivation of hydrodynamics. The standard
derivation is rather akin to the Ehrenfest theorem of elementary quantum
mechanics which shows that the averages of position and momentum operators obey
classical equations of motion. Yet a description of emergent classicality
must involve much more than that \cite{Har6}.
Firstly, it must demonstrate decoherence of the local
densities, thereby allowing us to talk about probabilities for their histories.
Secondly, it should not be restricted to a special initial state.
Whilst it is certainly plausible that many initial states will tend to
the local equilibrium state, the standard derivation
does not obviously apply to superpositions of macroscopic states, which are exactly the states
a description of emergent classicality is supposed to deal with.

This paper is part of a general programme, initiated in Refs.\cite{Hal2,Hal3,Hal4},
to obtain a more general derivation of hydrodynamic equations from the
underlying quantum theory, using the decoherent histories approach.
The aim in particular is to consider reasonably general classes of initial
states and to demonstrate decoherence of the local densities, without
appealing to environmentally-induced decoherence, and to show that
the probabilities for histories are peaked about equations of motion
of the hydrodynamic type. In this paper the particular system we will
apply the programme to is a chain of linearly coupled oscillators.

The general sketch of the programme, which we will work out in detail
in this paper, is as follows. We start from the simple observation
that exactly conserved quantities define an exactly decoherent
set of histories, essentially because the projectors in the decoherence
functional commute with the Hamiltonian \cite{HLM}. It is therefore
expected that the histories will remain approximately decoherent
as we go from $k = 0$ to non-zero values of $k$ in the local densities
$ n( \k) $, ${\bf g} (\k )$, $ h( \k )$.
In Ref.\cite{Hal2} it was shown that a useful way to organize this idea
is to decompose the initial state of the system into a superposition of
states, $ | n, {\bf g}, h \ra$, which are approximate
eigenstates of the local densities. It is then very plausible (and verifiable
in specific models) that
such states remain approximate eigenstates of the local densities
under time evolution, for sufficiently small $k$ (since it is clearly
exactly true in the limit $k \ria 0 $). Here, ``sufficiently small''
means that $k^{-1}$ should be much greater than the correlation length
of each local density eigenstate. The preservation in time of these
states means that histories of them will be approximately decoherent.

Given decoherence we may then look at the probabilities for histories.
Decoherence also indicates that each element of the superposition
of local density eigenstates may be treated separately. We
therefore consider the probabilities for histories of local
densities with the local density eigenstate as the initial state.
For sufficiently coarse-grained histories the probabilities for
the local densities are strongly peaked at each time about the
average value of the local densities, averaged in the local
density eigenstate $ | n,{\bf g},h \ra $. Since the local
densities are sums of one-particle operators (to lowest order in
interactions), this is the same as averaging in the one-particle
reduced density operator $\rho_1$ constructed by tracing $ | n,
{\bf g}, h \ra \la  n, {\bf g}, h |$ over all but one particle states.
The density operator
$\rho_1$ is clearly not the same as the (one-particle version of the)
local equilibrium state
(\ref{1.10}), although the two states are clearly very similar,
since they are both very localized in the local densities. Hence,
to complete the derivation of the hydrodynamic equations, it is
necessary to show that $\rho_1$ tends to the local equilibrium
state after a period of time. This is clearly extremely plausible on
physical grounds and may be proved in explicit cases, as in this
paper. (And indeed, this is much weaker than asserting that
{\it any} initial state tends to the local equilibrium state.)

In brief, the whole story works in particular models contingent only
on constructing local density eigenstates and showing that they
have the desired properties: that they are preserved in form
under time evolution for sufficiently small $k$, and that
they are effectively equivalent to the local equilibrium distribution
after a period of time. The point of this paper is to show this
for the linear oscillator chain.

The detailed connection between conservation and decoherence is discussed in Section 2,
as is the construction of approximate eigenstates of the local densities.

In Section 3, we describe the dynamics of the linear oscillator chain.
We consider two types of chain: the simple chain, where only neighbouring
particles are coupled, and the bound chain, where each particle is
in addition bound to an origin by a harmonic potential.
We consider both finite and infinite chains.
The most important results are the correlation functions, which establishes
the scale on which coarse-graining is required for decoherence.

In Section 4, as a preparation for proving decoherence of the local densities
of the chain of oscillators, we consider a simplified set of variables,
namely, the total momentum contained in a subsection of the chain.
We show that the eigenstates of this quantity remain approximate
eigenstates under time evolution as long as the size of the chain subsection
is much greater than the correlation length.

In Section 5, we consider the local densities of the chain. We prove
that approximate eigenstates remain
approximate eigenstates, for $k^{-1}$ much larger than the correlation
length of the chain.

In Section 6, we consider the probabilities for histories.
In the case of a finite simple chain, we show that the averages of
number and momentum density obey a closed set of equations (although
there is no evolution to local equilibrium in this case).
For the infinite bound chain, we show that the density operator
$\rho_1$ does indeed tend to the local equilibrium state and we derive
the resultant hydrodynamic equations.

We discuss our results in Section 7.

The idea that local densities should define a natural set of decoherent
histories as a result of their approximate conservation was first put forward
by Gell-Mann and Hartle \cite{GH2}. This idea, and the related possibility
of deriving hydrodynamic equations, has been developed by numerous
authors \cite{BrH,Hal2,Hal3,Hal4,Ana,CaH1,CaH2}.
This work is perhaps most closely related to that of Brun and Hartle \cite{BrH},
who analyze the linear oscillator chain using the decoherent histories approach.
Their approach was rather different in that they considered coarse-grainings
in which the centre of mass coordinates of chain subsections were specified,
rather than the local densities considered here, and they evaluated the
decoherence functional explicitly, rather than examine the evolution of
eigenstates of the variables of interest.

\section{Decoherence and Conservation}

We begin by describing the connection between decoherence and conservation,
It is well-known that exactly conserved quantities are exactly
decoherent \cite{HLM}. The simple reason for this is that the projectors
commute with the unitary evolution operator. The projectors $
P_{\a_k}$ on one side of the decoherence functional (\ref{1.1}) may therefore
be brought up against the projectors $P_{\a_k'} $ on the other
side, hence the decoherence functional is diagonal.
(In the situation considered here, in which there are three conserved
quantities involved, these quantities must in addition commute
with each other, but this is clearly the case.)

We would like to extend this idea to approximate decoherence in
the case of approximate conservation. It turns out that the above
argument is better formulated in a different way for the purposes
of generalization \cite{Hal2}. Suppose the initial state is pure and
consider the decoherence functional,
\begin{equation}
D(\au, \au') = {\rm Tr} \left( P_{\a_n} U_{n-1,n} \cdots P_{\a_2} U_{12}
P_{\a_1} | \Psi \ra \la \Psi | P_{\a_1'} U_{12}^{\dag} P_{\a_2'} \cdots
U_{n-1,n}^{\dag} \right)
\label{2.0}
\end{equation}
where $U_{12}$ is the usual unitary evolution operator between
times $t_1 $ and $t_2$.
Suppose the histories are projections onto some conserved
quantity, $Q$. Now let the initial state be a superposition of
eigenstates of $Q$,
\begin{equation}
| \Psi \ra = { 1 \over \sqrt{2} } \left( | Q_1 \ra + |Q_2 \ra
\right)
\end{equation}
where
\begin{equation}
Q | Q_a \ra = Q_a | Q_a \ra
\label{2.1}
\end{equation}
and $a=1,2$. Since the $P_{\a}$'s are projections onto $Q$, we
have $P_{\a} | Q_a \ra = | Q_a \ra $, if $\a$ is suitably chosen,
otherwise we get $ P_{\a} | Q_a \ra = 0$. Hence the only non-zero
off-diagonal terms of the
decoherence functional are of the form,
\begin{equation}
D(\au, \au') = \half {\rm Tr} \left( P_{\a_n} U_{n-1,n} \cdots P_{\a_2} U_{12} |
Q_1 \ra \la Q_2 | U_{12}^{\dag} P_{\a_2'} \cdots U_{n-1,n}^{\dag} \right)
\end{equation}
But $Q$ is conserved, hence $ U_{12} |Q_a \ra = | Q_a \ra $ and
\begin{equation}
D(\au, \au') = \half {\rm Tr} \left( P_{\a_n} U \cdots P_{\a_2} |
Q_1 \ra \la Q_2 | P_{\a_2'} \cdots U^{\dag} \right)
\end{equation}
Proceeding in this way to the end of the chain,
\begin{equation}
D(\au, \au') = \la Q_2 | Q_1 \ra = 0
\end{equation}
for all pairs of distinct histories $\au, \au' $.
Hence decoherence comes about because neither the projections nor
the unitary evolution disturb the states $ |Q_a \ra $, and hence
the two orthogonal states $ |Q_1\ra $, $ | Q_2 \ra $ are brought
together at the final time and overlapped to give zero.

Let us now suppose that we have some operator $Q$ such that under time
evolution, its eigenstates are mapped into approximate
eigenstates. That is, we initially have Eq.(\ref{2.1}), but under evolution
to time $t$,
\begin{equation}
Q (t) | Q_a \ra \approx \la Q (t) \ra | Q_a \ra
\end{equation}
(where the average on the right-hand side is in the state $ | Q_a \ra $).
More precisely, this can be expressed as
\begin{equation}
{ \left( \Delta Q (t) \right)^2 \over \la Q (t) \ra^2 } << 1
\label{2.2}
\end{equation}
where
\begin{equation}
\left( \Delta Q (t) \right)^2 = \la Q^2(t) \ra - \la Q(t) \ra^2
\end{equation}
Eq.(\ref{2.2}) means that the state remains strongly peaked in the variable $Q$
under time evolution. The states are then approximate eigenstates
of the projectors at each time as long as the widths of the projectors
are chosen to be much greater than $ ( \Delta Q (t) )^2$.
The same argument goes through although this time only approximately.
Approximate decoherence
is therefore assured for sufficiently coarse-grained histories of operators $Q$ and
superpositions of initial states each of which have the property
that they remain strongly peaked in $Q$ under time evolution (as
characterized by Eq.(\ref{2.2})).

A simple example is the case of the coherent states of the
harmonic oscillator. These states are preserved in form under time
evolution, hence will always be approximate eigenstates of
projections onto position, momentum, or phase space, provided that
the widths of the projections are chosen to be much greater than
the uncertainties in $p$ and $q$ in the coherent states.
In this example, there is no obvious local conservation law. For this
reason, it is perhaps more accurate to speak of approximate determinism,
rather than approximate conservation. So very broadly speaking,
approximate decoherence of histories will arise when there is an approximate
determinism in the underlying quantum theory.

Returning to the local densities, we require a set of
states $ |n, {\bf g}, h \ra $ which are eigenstates of
all three local densities. Since the local densities do not
commute with each other, except in the limit $k \ria 0 $,
we can only find states which are approximate eigenstates.
The number and momentum density are both operators of the form,
\begin{equation}
A = \sum_{n=1}^N A_n
\end{equation}
as is the local energy density, if
we ignore the interaction term. For such operators it follows that
\begin{equation}
( \Delta A)^2 = \sum_n (\Delta A_n)^2 + \sum_{n \ne m} \s
(A_n, A_m)
\label{2.3}
\end{equation} and
\begin{equation}
\la A \ra^2 = \sum_{n,m} \la A_n \ra \la
A_m \ra
\end{equation}
A state will be an approximate eigenstate of the
operator $A$ if
\begin{equation}
\frac { ( \Delta A)^2 } { \la A \ra ^2}  \
\ll \ 1
\end{equation}
The expression for $ \la A \ra^2 $ potentially
involves $N^2$ terms, as does the expression for $ ( \Delta A)^2
$, but the latter will involve only $N$ terms if the correlation
functions $ \s (A_n, A_m ) $ are very small or zero for $n \ne m $.
So simple product states will be approximate eigenstates
and will have $ (\Delta A)^2 / \la  A \ra^2 $
of order $ 1/N$. (See Refs.\cite{Hal2,Hal3} for more detailed examples
this argument).

Under time evolution, the interactions cause
correlations to develop. However, the states will remain
approximate eigenstates as long as the correlations are
sufficiently small that the second term in Eq.(\ref{2.3}) is much
smaller than order $N^2$. The interactions and the subsequent
correlations are clearly necessary in order to get interesting
dynamics and in particular the approach to local equilibrium. The
interesting questions is therefore whether there is a regime
where the effects of interactions are small enough to permit
decoherence but large enough to produce interesting dynamics.
The fact that the variables we are interested in are locally conserved
indicates that there is such a regime.
The important point is that the local
densities become arbitrarily close to exactly conserved quantities as $k \ria 0$.
This means that, at any time, $ ( \Delta A)^2 / \la A \ra^2 $ becomes arbitrarily
close to its initial value (which is of order $1/N$) for sufficiently small $k$.

In the examples we look at it in the following sections, we will see that
an uncorrelated initial state develops correlations with a typical lengthscale
(or extending to a certain number of particles down the chain). These correlations
typically then decay with time. What we will
find is that the second term in Eq.(\ref{2.3}) will remain small as long
as $k^{-1}$ is much greater than the correlation length. Hence the key physical
aspect is the locality of the interactions, meaning that only limited local correlations
develop, together with the coarse-graining scale $k^{-1}$ which may be chosen
to be sufficiently large that the correlation scale is not seen.
Differently put, as $k$ increases from zero, departing from exact decoherence,
it introduces a lengthscale $k^{-1}$. Since the decoherence functional is a dimensionless
quantity, clearly nothing significant can happen until $k^{-1}$ becomes comparable
with another lengthscale in the system. The natural scale is the correlation
length in the local density eigenstates.

The scheme described here would be executed most transparently if
we used states which become exact eigenstates of the conserved
quantities in the limit $k \ria 0$, thereby always maintaining the
closest connection with exact conservation. In the next section we
will in fact use Gaussians as the approximate eigenstates, because
they are the easiest states to work with. These will not be exact
eigenstates of the exactly conserved quantities in the $k \ria 0 $
limit, although this not in fact matter very much, for reasons
outlined above. (Furthemore, the decoherence functional is always
exactly diagonal for any initial state in the $k \ria 0 $ limit,
for the reasons stated at the beginning of this section, but we do
not need to exploit this here.)

\section{Chains of Oscillators}

In this and the following sections, we show how the general
programme outlined above may be worked out in detail in linear
oscillator models. These have the advantage that they can be
solved exactly. In particular, the time-development of the
correlation functions and eigenstates of the local densities can
be computed reasonably explicitly.

\subsection{The Models and Their Classical Solutions}

We consider a chain of point particles which are coupled to each
other by a nearest-neighbour linear interaction. We also allow the
possibility that each particle is harmonically bound to one of a
series evenly distributed points, separated by distance $b$, say.
The Hamiltonian is
\begin{equation}
H = \sum_{n=1}^N \left[ { p_n^2 \over 2m} + {\nu^2 \over 2} ( q_n
- q_{n-1} )^2 + { K \over 2} ( q_n - b_n )^2 \right]
\end{equation}
where $b_n = n b $. We will consider the two cases $K=0$ (the
simple chain) and $K \ne 0 $ (the harmonically bound chain). In
the bound chain case, it is also useful to consider the case $b_n
= 0$, which corresponds to the situation in which the whole chain
moves in a harmonic potential. (In fact, for the classical
solutions, $b_n$ is readily absorbed into $q_n$, but this makes a
difference to the local densities considered below). We initially
consider a finite number $N$ of particles but we also consider the
case of $N$ infinite.

The equations of motion are
\begin{equation}
m \ddot q_n + K (q_n - b_n)  = \nu^2 ( q_{n+1} - 2 q_n + q_{n-1} )
\end{equation}
where we take $ q_{N+1} = q_1 $. This system has been discussed
and solved in many places. A particularly useful reference for the
case of an infinite chain is the treatment by Huerta and Robertson
\cite{HuR1,HuR2}. (See also Refs.\cite{Fey,Thi,Aga,TeS}). The
solution may be found by introducing the normal modes, $Q_{\a}$,
\begin{equation}
q_n = b_n + \sum_{\a =1}^N \ { e^{ 2 \pi i \a n / N } \over
N^{\half} } Q_{\a}
\label{3.3}
\end{equation}
which obey
\begin{equation}
\ddot Q_{\a} + \omega^2_{\a} Q_{\a} = 0
\end{equation}
where
\begin{equation}
\omega_{\a} = \left( \frac {K} {m}  + \frac {4 \nu^2}{m}
\sin^2 \left( { \pi \a \over N } \right) \right)^{\half}
\label{3.5}
\end{equation}
The solution may be written,
\begin{equation}
q_n (t) = b_n + \sum_{r=1}^N \left[ f_{r-n} (t) q_r (0) + {
g_{r-n} (t) \over m \Omega } p_r (0) \right]
\end{equation}
where
\begin{eqnarray}
f_n (t) &=& \frac {1} {N} \sum_{\a=1}^N e^{2 \pi i \a n / N} \ \cos
\left( \omega_{\a} t \right)
\\
\frac {g_n (t)} {\Omega} &=& \frac {1} {N} \sum_{\a=1}^N e^{2 \pi
i \a n / N} \ \frac {\sin \left( \omega_{\a} t \right)} {\omega_{\a} }
\end{eqnarray}
Here, $\Omega^2 = ( K + 2 \nu^2) / m $. The solution for $p_n (t) $ is given by
\begin{equation}
p_n (t) = m \dot q_n (t)
\end{equation}

In the limit of an infinite number
of particles the solution is
\begin{equation}
q_n (t) = b_n + \sum_{r=-\infty}^\infty \left[ f_{r-n} (t) q_r (0)
+ { g_{r-n} (t) \over m \Omega } p_r (0) \right]
\label{3.8}
\end{equation}
For the simple chain, $K = 0$, the solution is then given in terms of Bessel
functions \cite{HuR1}, and we have
\begin{equation}
f_r(t) = J_{2r} (2 \omega t)
\end{equation}
where $ \omega^2 = \nu^2 / m $, and
\begin{equation}
g_r(t)  = \Omega \int_0^t dt' \ J_{2r} (2 \omega t')
\end{equation}
The appearance of the time integral in the expression for $g_r (t)$ is in fact related
to the motion of the centre of mass of the whole chain, which is not of interest for our
considerations. (We imagine the whole system is contained somehow so that there is
no wavepacket spreading).
It also somewhat obscures the discussion of correlations, which is the main
thing we are interested in. The relevant behaviour in $g_r (t)$ is best exhibited in
terms of the difference variables, $g_{r+1} - g_r$. Using a simple recurrence relation for the
Bessel functions, these are given by
\begin{equation}
g_{r+1} (t) - g_r (t) = - 2 \Omega J_{2r+1} (2 \omega t )
\label{3.11}
\end{equation}
We will discuss this in more detail below.

For the bound chain, $ K \ne 0 $, it is most useful to work in the
regime in which the interaction between particles is much weaker
than the binding to their origins, so $ \nu^2  << K $. In this
case, the solution then is \cite{HuR1},
\begin{equation}
f_r (t) \approx J _r (\gamma \Omega t ) \cos \left( \Omega t - \pi r / 2 \right)
\end{equation}
and
\begin{equation}
g_r (t) \approx J_r ( \gamma \Omega t ) \sin \left( \Omega t - \pi r / 2 \right)
\end{equation}
where $\gamma = ( \omega / \Omega)^2 $, so $\gamma << 1 $.

The general behaviour of the solutions in both cases is easily
seen. The functions $f_{r-n}(t)$ and $g_{r-n} (t)$ loosely
represent the manner in which an initial disturbance of particle
$r$ affects particle $n$ after a time $t$, and is given in both
the bound and unbound case by the properties of Bessel functions.
Recall the following forms of the Bessel functions \cite{AbSt}. For small
arguments we have
\begin{equation}
J_n (x) = \left( {x \over 2 } \right)^n \left( \frac {1} {\Gamma (n+1) } -
\frac { (x/2)^2 } { 2! \Gamma (n+3) } + \cdots \right)
\label{3.b1}
\end{equation}
(This is for $n>0$. For $n<0$ we use $J_{-n} (x ) = (-1)^n J_n (x) $).
For large arguments we have the asymptotic form
\begin{equation}
J_n (x) \sim \left( \frac {2} {\pi x} \right)^{1/2} \ \cos \left(
x - \pi n / 2 - \pi/ 4 \right)
\label{3.15}
\end{equation}
Hence the Bessel functions start out at zero (except for $n=0$, where $J_0 (0) = 1$),
grow exponentially, and then go into a slowly decaying oscillation.
For large $n$ and fixed $x$ we have
\begin{equation}
J_n (x) \sim ( 2 \pi n)^{- \half} \left( \frac { e x} {2 n} \right)^n
\end{equation}

A point not immediately obvious from these standard asymptotic
forms, and which will turn out to be important, is that the
different Bessel functions each go into their oscillatory regions
at different values of $x$. In particular, one can estimate from
the plots of the Bessel functions that $J_n (x)$ goes into its
oscillatory regime when $ x $ is of order $n$, by which stage
$J_n(x)$ is therefore of order $ n^{-1/2}$. In terms of the
behaviour of the chain, this means that distant pairs of particles
never come to influence each other very much, even after long
periods of time: at short times, particles separated by $n$ have
exponentially suppressed correlation, like $x^n / n! $ and at long
times, their correlations are also suppressed, like $n^{-1/2}$.
This particular aspect turns out to be crucial for our purposes.

Another important observation is that in the oscillatory regime,
the Bessel function $J_n (x)$ has only a very limited dependence on $n$, namely it has
the form (\ref{3.15}) for some $n$, plus the three possible phase shifts of $\pi/2$.
The significance of this for the chain is that when the functions $ f_{r-n} (t)$ and
$ g_{r-n}(t)$ have entered the oscillatory regime, the conditions at particles
$r$ and $n$ and everywhere in between are approximately the same. This
feature is clearly relevant to the approach to local equilibrium.

\subsection{Correlation Functions }

As described in Section 2, we are interested in the time-development
of eigenstates of the local densities, and this boils down to the behaviour of
the various correlation functions of the system.
We define the correlation function
\begin{equation}
\s (A, B ) = \half \langle A B + B A \rangle - \la A \ra \la B \ra
\end{equation}
Because the system is linear the classical solutions described
above may be used to discuss the solutions in the Heisenberg
picture in the quantum case. In fact, the only quantum calculations
in this paper have essentially been done already in Section 2,
and the remaining analysis is essentially
the same as for a classical stochastic system.

For simplicity, we will concentrate on Gaussian initial states (which will
of course remain Gaussian, because they system is linear), although these
will be sufficient for our purposes. The variances of these Gaussians are
restricted by the requirement that the state is pure. (They are also
of course restricted by the uncertainty principle).
We will consider two different types of Gaussian initial states which
can be approximate eigenstates of the local densities. The first type we consider
are product states, so have no initial correlations between different particles.
The second type are the coherent states of the normal modes, and
are naturally suggested by the normal mode decomposition Eq.(\ref{3.3}).
We will see in the next section that these are eigenstates of the local
densities as long as the correlation functions remain sufficiently
small.

\subsubsection{Normal Mode Coherent States}

Taking the second type first, we therefore consider the set of Gaussian states
which have
\begin{eqnarray}
\s (Q_{\a}, Q^*_{\b}) &=& \frac {\hbar } {2 m \omega_{\a}}  \delta_{\a \b}
\\
\s( K_{\a}, K^*_{\b}) &=&  \half \hbar m \omega_{\a}\delta_{\a \b}
\\
\s (Q_{\a}, K_{\b} ) &=& 0
\end{eqnarray}
where $K_{\a}$ is the momentum conjugate to $Q_{\a}$ and note that
$Q^*_{\a} = Q_{-\a} $. Because these are the coherent states of the
harmonic oscillator, these correlation functions all remain of this
form under time evolution, and the only time-development of the states
is in terms of their centres, $Q_{\a} (t) $, $K_{\a} (t)$, which follow
the classical equations of motion.
For the case of the simple chain, $K=0$, we have $\omega_{\a} = 0$ when
$\a = N $. This corresponds to the centre of mass of the whole
chain, and may be quite simply omitted.

In terms of the original coordinates $q_n$ and $p_n$, we have the
correlation functions
\begin{eqnarray}
\s(q_n, q_m ) &=& \frac {1} {N} \sum_{\a =1}^N  \ \frac {\hbar } {2 m \omega_{\a}} \ e^{2 \pi i \a (n-m)/N }
\label{3.20}
\\
\s (p_n, p_m) &=& \frac {1} {N} \sum_{\a=1}^N   \ \half \hbar m \omega_{\a} \ e^{2 \pi i \a (n-m)/N }
\\
\s (q_n, p_m ) &=& 0
\end{eqnarray}
These correlation functions are constant in time, and this feature makes this
case a useful one to study. Also notice that $( \Delta q_n)^2 $ and
$ ( \Delta p_n )^2 $ are independent of $n$. The correlation functions will typically
decay very rapidly with increasing $ | n - m | $, since they are sums of rapidly
oscillating terms. This is especially true in the case $K \ne 0 $ with $K >> \nu^2$,
since then $\omega_{\a}$, Eq.(\ref{3.5}), is a constant to leading order. (We will
a similarly effect in more detail in the next subsection). In the case $K = 0$, we have
\begin{equation}
\omega_{\a} =  \frac {2 \nu} {m^{\half}} \sin \left( \frac{ \pi \a } {N} \right)
\end{equation}
The correlation function $\s (p_n, p_m) $ clearly still decays for large
$ | n - m | $, but this is less obvious for $ \s (q_n, q_m) $, Eq.(\ref{3.20}),
because the denominator becomes very small close to $\a = N $. However, if we
eliminate by hand a small cluster of modes close to $\a = N$, we get satisfactory
decay properties for this correlation function, and we will assume that this has been
done.

\subsubsection{Uncorrelated Initial States for the Infinite Chain}

We now consider uncorrelated initial states for the infinite
chain, so the correlation functions $ \s
(q_n, q_m) $, $\s (q_n, p_m) $ and $ \s (p_n, p_m) $ all vanish at
the initial time for $n \ne m $. We are then interested in the
behaviour of these three types of correlation functions at later
times.

From the solution Eq.(\ref{3.8}), with the assumption of no initial correlation between
the particles, we have
\begin{eqnarray}
\s (q_n (t), q_m (t) ) =
\sum_r \left( f_{r-n} (t) f_{r-m} (t) (\Delta q_r )^2
+  \frac {1} {m^2 \Omega^2} g_{r-n} (t) g_{r-m} (t) ( \Delta p_r)^2 \right.
\nonumber \\
\left. +  \frac {1} {m \Omega} \left[ f_{r-n} (t) g_{r-m} (t)
+  f_{r-m} (t) g_{r-n} (t) \right] \s ( q_r, p_r )
\right)
\label{3.24}
\end{eqnarray}
and similarly for $ \s (q_n (t), p_m (t) ) $ and $ \s ( p_n (t), p_m (t) )$.
These expressions simplify if we make the further assumption that
the initial values of $ (\Delta q_r)^2 $, $(\Delta p_r)^2 $ and $ \s (q_r, p_r) $
are independent of $r$ (we will show below how to go beyond this assumption).
In this case we obtain
\begin{eqnarray}
\s (q_n (t), q_m (t) ) &=& a_{nm} (t) (\Delta q)^2 + 2 e_{nm} (t) \s (q,p) + d_{nm} (t) (\Delta p)^2
\label{3.25}
\\
\s (q_n (t), p_m (t) ) &=& b_{nm} (t) (\Delta q)^2 + [ a_{nm} (t) + k_{nm} (t) ] \s (q,p) + e_{nm} (t) (\Delta p)^2
\label{3.26}
\\
\s (p_n (t), p_m (t) ) &=& c_{nm} (t) (\Delta q)^2 + 2 b_{nm} (t) \s (q,p) + a_{nm} (t) (\Delta p)^2
\label{3.27}
\end{eqnarray}
where
\begin{eqnarray}
a_{nm} (t) &=& \sum_r f_{r-n} (t) f_{r-m} (t)
\\
b_{nm} (t) &=& m \sum_r f_{r-n} (t) \dot f_{r-m} (t)
\\
c_{nm} (t) &=& m^2 \sum_r \dot f_{r-n} (t) \dot f_{r-m} (t)
\\
d_{nm} (t) &=& \frac {1} {m^2 \Omega^2} \sum_r g_{r-n} (t) g_{r-m} (t)
\\
e_{nm} (t) &=& \frac {1} {m \Omega^2} \sum_r g_{r-n} (t) \dot g_{r-m} (t)
\\
k_{nm} (t) &=& \frac {1} {\Omega} \sum_r \dot f_{r-n} (t) g_{r-n} (t)
\end{eqnarray}
Since the coefficients $f_n (t)$ and $g_n (t)$ are all given by Bessel functions,
these expressions can be evaluated using the following Bessel function addition theorem:
\begin{equation}
J_n ( 2x ) = \sum_{k=- \infty}^{\infty} J_{n-k} (x) J_k (x)
\label{3.34}
\end{equation}

For the bound chain, $K \ne 0$, the coefficients are \cite{HuR1},
\begin{eqnarray}
2 a_{nm} (t) &=& \delta_{nm} + J_{n-m} ( 2 \gamma \Omega t) \cos \left[ 2 \Omega t
-\half (n-m) \pi \right]
\\
2b_{nm} (t) &=& - m \Omega J_{n-m} ( 2 \gamma \Omega t) \sin \left[ 2 \Omega t
-\half (n-m) \pi \right]
\\
2 c_{nm} (t) &=& (m \Omega)^2 \delta_{nm} - (m \Omega)^2 J_{n-m} ( 2 \gamma \Omega t) \cos \left[ 2 \Omega t
-\half (n-m) \pi \right]
\\
d_{nm} (t) &=& (m \Omega)^{-4} c_{nm} (t)
\\
e_{nm} (t) &=& - (m \Omega)^{-2} b_{nm} (t)
\\
k_{nm} (t) &=& - (m \Omega)^{-2} c_{nm} (t)
\end{eqnarray}
All of these coefficients, and hence all of the correlation
functions, decay exponentially for large $ | n - m | $.
Furthermore, in the limit $ t \ria \infty $ we have
\begin{eqnarray}
\s (q_n (t), q_m (t) ) & \ria & \half \delta_{nm} \left(   (\Delta q)^2
+ \frac {1} { m^2 \Omega^2} (\Delta p)^2 \right)
\\
\s (q_n (t), p_m (t) ) & \ria & 0
\\
\s (p_n (t), p_m (t) ) & \ria & \half \delta_{nm} \left(  m^2 \Omega^2  (\Delta q)^2 +
(\Delta p)^2 \right)
\end{eqnarray}
We also find that $ \la q_n (t) \ra  \ria b_n $ and $\la p_n(t) \ra \ria 0$
as $ t \ria \infty$.
We thus obtain an equilibrium distribution, since the corresponding phase
space distribution function for the whole chain is
\begin{equation}
w = \prod_n \exp \left( - \frac {1} {k T} \left[
\frac {p_n^2} {2m} + \half m \Omega^2 (q_n - b_n)^2 \right] \right)
\end{equation}
where we identify the temperature as
\begin{equation}
k T = \frac {1} {2m} \left(  m^2 \Omega^2  (\Delta q)^2 +
(\Delta p)^2 \right)
\end{equation}

For the simple chain, $K = 0$, we have \cite{HuR1},
\begin{eqnarray}
2 a_{nm} (t) &=&  \delta_{nm} + J_{2n-2m} (4 \omega t )
\label{3.38}
\\
2 b_{nm}(t) &=& m \omega \left[ J_{2n-2m-1} (4 \omega t) - J_{2n-2m+1} (4 \omega t)
\right]
\\
2c_{nm} (t) &=& (m \omega)^2 \left[ J_{2n-2m+2} (4 \omega t) + J_{2n-2m-2} (4 \omega t)
- 2 J_{2n-2m} (  4 \omega t) \right.
\label{3.39}
\\
& & \left. -\delta_{n,m-1} - \delta_{n,m+1} + 2 \delta_{nm} \right]
\\
d_{nm} (t) &=& \frac{t} {2 \omega m^2} \left[ \int_0^{4 \omega t} J_0 (y) dy - J_1 (4 \omega t)
\right] - \frac{1} {(2 \omega m)^2} \sum_{j=1}^{|n-m|} \int_0^{4 \omega t} J_{2j-1} (y) dy
\\
e_{nm} (t) &=& \frac{1} {4 \omega m} \int_0^{4 \omega t} J_{2n-2m} (y) dy
\end{eqnarray}
(The explicit form of $k_{nm} (t)$ is not required).
The first three of these coefficients, like the bound case, are exponentially
suppressed for large $ |n - m |$. This means that the behaviour of the correlation
$ \s (p_n (t), p_m (t) ) $, which depends only on these three coefficients,
has the expected behaviour, but the other correlation functions do not have
this property. In particular, from the behaviour of $ \s (q_n(t), q_m(t))$,
the whole chain appears to become highly correlated.
In the long-time limit, we find
\begin{eqnarray}
\s (q_n (t), q_m (t) ) & \ria & \half \delta_{nm} (\Delta q)^2 + \frac{t} {2 \omega m^2}
( \Delta p)^2
\\
\s (q_n (t), p_m (t) ) & \ria & \frac {1} {4 \omega m} ( \Delta p)^2
\\
\s (p_n (t), p_m (t) ) & \ria & \half (m \omega)^2 \left[
2 \delta_{nm}-\delta_{n,m-1} - \delta_{n,m+1} \right] ( \Delta q)^2
+ \half \delta_{nm} ( \Delta p)^2
\end{eqnarray}
(where for simplicity we have taken $\s (q_n,p_n) = 0 $ initially).
This is not an equilibrium distribution, and the behaviour of
$(\Delta q_n(t))^2 $ at late times indicates diffusive behaviour.

This growth of correlations and variances is essentially
unphysical, since any realistic system is contained in some way,
so the spreading cannot proceed beyond the size of the container
(and indeed there is no such effect in the $K \ne 0 $ case). It
is, however, difficult to factor out this unphysical effect in a
convenient way. The long-time limits of Ref.\cite{HuR1}, which we have
followed closely, are dominated by this spreading effect in the $K
= 0 $ case. In an attempt to understand this, the authors of
Ref.\cite{HuR1} considered a different set of models in Ref.\cite{HuR2}, in which
the chain was fixed at one end. This avoided the diffusive growth
encountered above, but still led to significant correlations
between all particles on the chain in the long-time limit. This in
turn spoils the desired behaviour of the local densities,
discussed below. The upshot of this is that it is not possible
to prove decoherence of the local densities in the case of
the infinite chain with $K = 0$.

There are two simple ways in which the above results are easily generalized.
First of all, note that although we are focusing on Gaussian initial states,
expressions for the correlation functions such as Eq.(\ref{3.24}) are in fact valid
for {\it any} initial state, because of the linearity of the dynamics.

Secondly, in deriving Eqs.(\ref{3.25})--(\ref{3.27}), we assumed that the initial
variances are independent of $r$. This assumption was necessary in order to
be able to apply the Bessel function addition theorem Eq.(\ref{3.34}), and thereby
obtain explicit expressions for the coefficients, $a_{nm}$, $b_{nm}$ etc.
This is too restrictive, since it means that certain hydrodynamic variables
(such as temperature, which depends on $ (\Delta p_r)^2 $), are obliged to be
constant along the chain.
One can see, however, that these results easily extend to the case in which
the initial variances $ (\Delta q_r)^2$, $(\Delta p_r)^2$ and $ \s (q_r, p_r)$ have
a slow dependence on $r$ along the chain. The point is that because of the
decay of the functions $ f_{n-r}$, $g_{n-r}$ for large $ | n - r | $,
the sum over $r$ in Eq.(\ref{3.24}) is effectively restricted to a finite range,
namely, over the (actually quite small) range for which significant correlations exist.
As long as the initial variances vary significantly with $r$ only on a range larger
than the correlation range, then the calculation of correlation functions is
effectively equivalent to the case in which the variances are completely independent of
$r$. This means that in place of Eq.(\ref{3.25}), for example, we get the more general
result
\begin{equation}
\s (q_n (t), q_m (t) )  \approx  a_{nm} (t) (\Delta q_r)^2 + 2 e_{nm} (t) \s (q_r,p_r) + d_{nm} (t) (\Delta p_r)^2
\end{equation}
where $r$ on the right-hand side is taken to be mid-way between $n$ and $m$, and for
$n=m$ we get
\begin{equation}
(\Delta q_n(t) )^2 \approx a_{nn} (t) (\Delta q_n)^2 + 2 e_{nn} (t) \s (q_n,p_n) + d_{nn} (t) (\Delta p_n)^2
\label{3.40}
\end{equation}
This simple observation is important for obtaining interesting hydrodynamic equations
because it allows for the possibility of the system tending towards {\it local}
equilibrium, rather than equilibrium of the whole chain.

\section{Coarse Graining by Chain Subsections}

Although we are ultimately interested the local densities for the
chain variables, we will first consider some simpler variables whose
analysis is highly instructive. Namely, we take the variables of
interest to be the total momentum in a subsection of the chain
containing $M$ particles, so we define
\begin{equation}
P_M = \sum_{n=1}^M p_n
\end{equation}
where $M \ll N $. This is not quite the same as a locally conserved quantity,
but it is very similar, since, for the simple chain, the total momentum
is conserved. Therefore $P_M$ is an exactly conserved quantity when
$M= N$, and otherwise we might expect it to be approximately conserved.

As outlined in Section 2, to show that these variables are approximately
decoherent, we need to show that initial states satisfying the condition
\begin{equation}
\frac { ( \Delta P_M)^2 } { \la P_M \ra^2 } \ \ll 1
\label{4.2}
\end{equation}
will continue to satisfy it under time evolution, hence the initial state
remains an approximate eigenstate. We have
\begin{equation}
( \Delta P_M)^2 = \sum_{n=1}^M \sum_{m=1}^M \ \s ( p_n, p_m)
\end{equation}
and
\begin{equation}
\la P_M \ra^2 = \sum_{n,m}^M \la p_n \ra \la p_m \ra
\end{equation}
We take the case in which the initial correlations are zero and the
initial variances are the same all along the chain,
and we also take the initial value of $\s (q,p) $ to be zero. We thus obtain
\begin{equation}
( \Delta P_M (t) )^2 = C_M  (\Delta q)^2 +  A_M (\Delta p)^2
\end{equation}
where
\begin{eqnarray}
A_M (t) &=& \sum_{n=1}^M \sum_{m=1}^M a_{nm} (t)
\\
C_M (t) &=& \sum_{n=1}^M \sum_{m=1}^M c_{nm} (t)
\end{eqnarray}
and the coefficients $a_{nm}(t)$ and $c_{nm} (t)$ are given by
Eqs.(\ref{3.38}),(\ref{3.39}).
The two terms on the right are very similar, so for simplicity will will
concentrate on the second one. (Note also that these terms do not suffer
from the spreading problem discussed in the previous section. This is an advantage
of working with momenta, rather than positions).
From the expression (\ref{3.38}), we have
\begin{equation}
A_M (t) = \half \sum_{n=1}^M \sum_{m=1}^M J_{2n -2m} ( 4 \omega t )
\end{equation}
(where we have assumed that $N$ is sufficiently large that it is
effectively equivalent to the $N \ria \infty $ case).
Our aim is now to show that $A_M \ll M^2$ for all times,
for then the condition Eq.(\ref{4.2}) will be satisfied.

The expression for $A_M$ cannot be evaluated exactly, but its properties
may be seen from the asymptotic forms of the Bessel functions (\ref{3.b1}),(\ref{3.15}).
For small times, when the Bessel functions are all in the exponential
regime, $J_{2n -2m} $ is exponentially suppressed for large $ | n - m |$.
For larger times, the Bessel functions start to go into their oscillatory
form, where they are already small. Furthermore, because in the oscillatory
regime they depend on $n-m$ only through a simple phase, most of the terms
in the sum over $n$ and $m$ cancel out. Proceeding along these lines one
can see that $A_M$ will not come anywhere close to $M^2$ except for
small values of $M$. These features are easily confirmed by plotting
$A_M (t) $ for different values of $M$. For example, with $M = 5$,
$A_M (t)$ quickly decays to a value of about $0.1$, clearly much less
than $M^2 = 25$, thereafter going into a
slowly decaying oscillation.

As we shall see, the variances of the local densities differ from the momentum
of a chain subsection in that they are more complicated functions of the
correlation function, for Gaussian states, but the physical understanding
of their behaviour is essentially the same, which is why this simple example is
instructive.


\section{Decoherence of Local Densities}

We now come to the main point of this paper, which is to examiner
the behaviour of the local densities for the oscillator chain.
They are
\begin{eqnarray}
n(x) &=&  \sum_{n=1}^N \delta (q_n - x )
\label{5.1}\\
g(x) &=& \sum_{n=1}^N p_n \delta (q_n - x )
\label{5.2}\\
h(x) &=& \sum_{n=1}^N \left[ { p_n^2 \over 2m} + {\nu^2 \over 2} (
q_n - q_{n-1} )^2 + \half K ( q_n - b_n )^2 \right] \delta (q_n -
x ) \label{5.3}
\end{eqnarray}
Again it will often be very useful to work with the Fourier
transformed local densities,
\begin{eqnarray}
n(k) &=&  \sum_{n=1}^N e^{ i k q_n }
\label{5.4}
\\
g(k) &=& \sum_{n=1}^N p_n e^{ i k q_n }
\label{5.5}
\\
h(k) &=&  \sum_{n=1}^N \left[ { p_n^2 \over 2m} + {\nu^2 \over 2}
( q_n - q_{n-1} )^2 + \half K ( q_n - b_n )^2 \right] e^{ i k q_n
} \label{5.6}
\end{eqnarray}
The local number and local energy density are locally conserved. The
local momentum density is locally conserved only for the case of
the simple chain, $K = 0$.

We consider states which are approximate eigenstates of the local densities.
We will use Gaussian initial states, and we expect that these will be
approximate eigenstates of the local densities if we choose the
correlation functions $ \s (q_n, q_m) $, $\s (q_n, p_m) $ and $ \s
(p_n, p_m) $ to be zero, or at least sufficiently small, for $n
\ne m $.

For the Gaussian
states we consider here, the computation of variances of the local
densities is facilitated by the identity,
\begin{eqnarray}
\la & \exp &  \left( i \sum_n \left[ \a_n (q_n - \la q_n \ra ) +
\b_n ( p_n - \la p_n \ra ) \right] \right) \ra
\\
&=& \exp \left( - \sum_{nj} \left[ \half \a_n \a_j \s (q_n, q_j) +
\a_n \b_j \s (q_n, p_j) + \half \b_n \b_j \s (p_n, p_j) \right]
\right)
\label{5.7}
\end{eqnarray}
All of the variances of interest will therefore be functions of the
three basic types of correlation functions
$ \s (q_n, q_m) $, $\s (q_n, p_m) $ and $ \s (p_n, p_m) $ discussed in Section 3,
and the physical discussion will in fact be very closely related to that
of the simple case discussed in Section 4. Actually, the formula
Eq.(\ref{5.7}) also holds to quadratic order in $\a_n$ and $\b_n$
for {\it any} state, so
the results derived below on the basis of Gaussian states will be
valid for arbitrary states for small $k$.

We consider first the Fourier transformed number density $n (k)$.
In a general Gaussian state, we have
\begin{equation}
\la n(k) \ra = \sum_{j=1}^N \la e^{i k q_j } \ra
= \sum_{j=1}^N \exp \left( { i k \la q_j \ra - \half k^2 (\Delta q_j)^2 } \right)
\end{equation}
and
\begin{eqnarray}
( \Delta n ( k ) )^2 &=& \la n^{\dag} (k) n(k) \ra - | \la n(k) \ra |^2
\nonumber \\
&=& \sum_{j=1}^N \sum_{n=1}^N \la e^{ i k
q_j } \ra \la e^{ - ik q_{n} } \ra \left( e^{ k^2 \s (q_j,
q_{n} )} -1 \right)
\end{eqnarray}
The latter is to be compared with
\begin{equation}
| \la n (k) \ra |^2 = \sum_{j=1}^N \sum_{n=1}^N \la e^{i k q_j } \ra
\la e^{ - i k q_{n} } \ra
\end{equation}
With an initially uncorrelated state we have $ \s (q_j, q_{n} ) = 0 $ for $ j \ne n$
and we see that
\begin{equation}
( \Delta n ( k ) )^2 = \sum_j  | \la e^{ i k q_j } \ra |^2 \left( e^{ k^2 (\Delta q_j)^2 } -1 \right)
\end{equation}
From this we expect that
\begin{equation}
( \Delta n ( k ) )^2  \ll | \la n (k) \ra |^2
\label{5.11}
\end{equation}
as long as $k^{-1}$ does not probe on scales that are too short
(compared to $ \Delta q_j$),
so the state is an approximate eigenstate.

When there are correlations, as will arise over time, we expect that the state will
still be an approximate eigenstate on lengthscales $k^{-1}$ which are much greater than
the lengthscale of correlation. As $k$ increases from zero we have
\begin{equation}
{ ( \Delta n ( k ) )^2 \over | \la n (k) \ra |^2 } \sim { k^2 (
\Delta X)^2 \over  N^2 }
\end{equation}
where $ X = \sum_j q_j$ (the centre of mass coordinate of the whole chain).
This will be very small as long as $k^{-1}$ is much larger than
the typical lengthscale of a single particle. $ ( \Delta n ( k ) )^2 $
starts to grow very rapidly with $k$, and Eq.(\ref{5.11}) is no longer valid,
when $k^{-1}$ becomes less than the correlation length indicated
by $\s (q_j, q_{n} ) $. Hence the state is strongly peaked
about the mean as long as the coarse graining lengthscale $k^{-1}$
remains much greater than the correlation length of the
local density eigenstates.

Consider now the local momentum density. We have
\begin{equation}
\la g(k) \ra = \sum_j \left( \la p_j \ra + i k \s (q_j, p_j) \right)
\exp  \left( { i k \la q_j \ra - \half k^2 (\Delta q_j)^2 } \right)
\end{equation}
and at some length, we find
\begin{equation}
( \Delta g(k) )^2 = \sum_{j n} \la e^{ik q_j} \ra \la e^{-ik q_{n} } \ra
\left( A_{j n} + B_{j n} + C_{j n} \right)
\end{equation}
where
\begin{eqnarray}
A_{j n} &=& \s (p_j, p_{n} ) e^{k^2 \s(q_j, q_n) }
\label{5.15}
\\
B_{j n} &=&
\left( \la p_n \ra - i k \s (q_{n} -q_j, p_{n} ) \right)
\left( \la p_j \ra + i k \s (q_{j} -q_{n}, p_j ) \right)
\left( e^{k^2 \s_(q_j, q_{n})} - 1 \right)
\label{5.16}
\\
C_{j n} &=& -ik \left( \la p_n \ra - i k \s (q_{n}, p_{n} ) \right)
-ik \left( \la p_j \ra + i k \s (q_j, p_j ) \right)
+k^2 \s (q_j, p_{n} ) \s (q_{n}, p_j )
\end{eqnarray}
We require $( \Delta g(k) )^2 $ to be small in comparison to
\begin{eqnarray}
| \la g(k) \ra |^2 =\sum_{j n} \la e^{i k q_j } \ra
\la e^{ - i k q_{n} } \ra
\left(\la p_j \ra \la p_{n} \ra
+ ik \la p_{n} \ra \s (q_j, p_j) \right.
\nonumber \\
\left. - ik \la p_j \ra \s (q_{n}, p_{n} )
+k^2 \s (q_j, p_j) \s (q_{n}, p_{n} ) \right)
\label{5.17}
\end{eqnarray}
Despite the complexity of these terms the interpretation is reasonably simple.
As $k \ria 0 $,
\begin{equation}
\frac {( \Delta g(k) )^2} {|\la g(k) \ra|^2} \ria \frac {( \Delta P)^2}
{ \la P \ra^2}
\end{equation}
where $P = \sum_j p_j $ is the total momentum, and it is easy to confirm that
this is small (typically order $1/N$) for the states we are using.
As $k$ increases from zero, ${( \Delta g(k) )^2}$ will grow, and will
potentially contain of order $N^2$ terms, the same as $ {|\la g(k) \ra|^2} $.
By inspecting the three terms Eqs.(\ref{5.15})--(\ref{5.17}), however,
one can see that each of them are prevented from generating order $N^2$
terms as long as, respectively, the correlation functions
$ \s (p_j, p_{n})$, $\s ( q_j, q_{n}) $ and $ \s (q_j, p_{n})$
are suppressed for $ j \ne n$. As in the case of the number density,
$ {( \Delta g(k) )^2} / {|\la g(k) \ra|^2} $ will start to grow appreciably
as $k^{-1}$ approaches the lengthscale indicated by $\s (q_j, q_{n})$.

Consider now the energy density. The computation of $ (\Delta h(k)))^2 $ is
rather complicated, but since we are considering Gaussian states, the final
conditions on the correlation functions will be very similar to those on
the variance of the momentum density considered above, so we will not
carry out the computation explicitly. Instead, we consider a simpler
special case. Take the case where each oscillator is fixed to an
origin at $b_j$, and suppose that the binding to it is so strong that
each particle is well-localized around $b_j$. Then the integral over
a volume of the energy density $ h(x)$ is then approximately
equivalent to taking a coarse-graining consisting of the energy
contained in a chain subsection, similar to Section 4. We therefore
consider the variable
\begin{equation}
h_{M} = \sum_{j=1}^M h_j
\end{equation}
where
\begin{equation}
h_j = \frac {p_j^2} {2m} + \half K q_j^2
\end{equation}
(neglecting the interaction term). Following the general discussion of Section 2,
we will have $ (\Delta h_M)^2 \ll \la h_M \ra^2 $ provided that
\begin{equation}
\sum_{j n} \s (h_j, h_{n}) \ll \la h_M \ra^2
\end{equation}
so the left-hand side must be much smaller than order $M^2$.
The correlation function $ \s (h_j, h_{n}) $ is constructed
from terms like $ \s (q_j^2, q_{n}^2 )$ and similar functions,
and we have
\begin{equation}
\s (q_j^2, q_{n}^2) = 2 \s ( q_j, q_{n} )^2
\end{equation}
(recalling that $\la q_j \ra \approx b_j $ since the particles are tightly
bound). The discussion is then very similar to the case of Section 4,
with essentially the same result, which is that the state is an approximate
eigenstate as long as $M \gg 1$.

We have now shown that Gaussian states will be approximate
eigenstates of the local densities for $k$ sufficiently small
compared to the correlation length determine by $\s (q_j,
q_{n})$, and provided also that the other correlation functions
$ \s (q_j, p_{n})$ and $ \s (p_j, p_{n} )$ are small for $j
\ne n$. (These results also hold for non-Gaussian states for small
$k$). The results of Section 2 show that all the correlation
functions (excluding the infinite chain in the $K=0$ case) have
the desired properties for all time. This proves the desired
result that initial eigenstates of the local densities
remain approximate eigenstates
under time evolution, for sufficiently small $k$.

\section{Hydrodynamic Equations for the Local Densities}

We have shown in a variety of circumstances that eigenstates of local densities
are approximately preserved in form under time evolution, on sufficiently coarse-grained
scales, and therefore superpositions of them define decoherent sets of histories.
We may now therefore look at the diagonal elements of the decoherence functional,
representing probabilities for histories of these variables. These probabilities are
peaked about the average values of the local densities, averaged in the approximate
eigenstates we have been considering. (This is reasonably obvious, but a more
detailed proof appears in the appendix of Ref.\cite{Hal3}.) We will now show
that these average values
obey hydrodynamic equations.

\subsection{Local Conservation Equations}

The local densities satisfy the local conservation
laws,
\begin{eqnarray}
\dot n(x) &=& - \frac {1} {m} \frac {\partial g} {\partial x}
\label{6.1} \\
\dot g (x) &=& - \frac {\partial \tau} {\partial x} - K x n(x) + K \sum_j b_j \delta (q_j - x)
\label{6.2} \\
\dot h(x) &=& - \frac {\partial j} {\partial x}
\label{6.3}
\end{eqnarray}
They are actually more usefully written in momentum space,
\begin{eqnarray}
\dot n(k) &=& \frac {ik} {m} g (k)
\label{6.4} \\
\dot g(k) &=& ik \tau (k)  - K \sum_j (q_j - b_j) e^{i k q_j}
\label{6.5} \\
\dot h(k) &=& i k j(k)
\label{6.6}
\end{eqnarray}
where the currents $ \tau (k)$ and $j(k)$ are given by
\begin{eqnarray}
\tau(k) &=& \sum_j \frac{ p_j^2} {m} e^{i k q_j} + \nu^2 \sum_j
q_j \frac{ \left( e^{ i k q_{j-1} } - 2 e^{ i k q_j} + e^{ i k
q_{j+1}} \right)} {ik}
\label{6.7} \\
j(k) &=& \sum_j \frac{p_j} {m} \left( \frac{p_j^2} {2 m} + \half K
(q_j-b_j)^2 + \half \nu^2 (q_j - q_{j-1})^2 \right) e^{ikq_j}
\\ &+& \nu^2 \sum_j \frac{p_j} {m} (q_{j+1}-q_j) \frac{\left(e^{ik q_j} - e^{ik q_{j+1}}\right)} {ik}
\label{6.8}
\end{eqnarray}
(These are clearly finite as $k \ria 0 $.)
These currents are rather complicated in configuration space, except in the case where
we neglect the interaction term, when they are given by
\begin{eqnarray}
\tau (x) &=& \sum_j \frac {p_j^2} {m} \delta (q_j - x)
\label{6.9} \\
j(x) &=& \sum_j \frac {p_j} {m} \left( \frac {p_j^2} {2m} + \half K (q_j -b_j)^2 \right)
\delta (q_j -x )
\label{6.10}
\end{eqnarray}

Equations (\ref{6.1})--(\ref{6.3}) do not in general form a closed system, so do not
lead to hydrodynamic equations. To get a closed set, it is necessary to average
these equations in a set of states depending on just three fields, thereby
obtaining three equations for three unknowns. In the standard approach to
deriving hydrodynamics, the local equilibrium state is chosen. We will discuss this
below in Section C, but first we consider the simpler and instructive case
of the normal mode coherent states.

\subsection{Hydrodynamic Equations in the Case of Normal Mode Coherent States}

We consider the simple chain, $K = 0$, in the normal mode coherent states.
As we have shown, these states are strongly peaked in the local densities
so such states define a decoherent set of histories. The correlations functions of these
states are constant in time, so we do not expect a settling down to local equilibrium.
However, it turns out that the averages of the local densities still obey a simple
set of hydrodynamic equations, and this case turns out to be particularly transparent
and instructive.
Because the only dynamics in this case is contained in the motion of the centres $\la q_j \ra$,
$\la p_j \ra $, we need consider only the local number and momentum densities,
not the energy density (which may in fact be calculated from the number and momentum
densities in this case). Closure of the averaged conservation equations
(\ref{6.4}), (\ref{6.5}) is obtained in this case because there are two equations and
the states depend on just two sets of quantities, the $\la q_j \ra$ and
the $\la p_j \ra $.

In a general Gaussian state we have
\begin{eqnarray}
\la n(k) \ra &=& \sum_j \exp \left( i k \la q_j\ra - \half k^2 (\Delta q_j)^2 \right)
\label{6.12a} \\
\la g(k) \ra &=& \sum_j \left[ \la p_j \ra + i k \s (q_j, p_j ) \right]
\exp \left( i k \la q_j\ra - \half k^2 (\Delta q_j)^2 \right)
\label{6.13a} \\
\la \tau (k) \ra &=& \sum_j  \left[ \frac {1} {m} \left( \la p_j
\ra + i k \s (q_j, p_j) \right)^2 + C \right] \exp \left( i k \la
q_j\ra - \half k^2 (\Delta q_j)^2 \right)
\nonumber \\
&+& \frac {\nu^2} {ik}  \sum_j  ( \la q_{j+1} \ra - 2 \la q_j \ra + \la q_{j-1} \ra)
\exp \left( i k \la
q_j\ra - \half k^2 (\Delta q_j)^2 \right)
\label{6.14a}
\end{eqnarray}
where
\begin{equation}
C_j  = \frac{1} {m} (\Delta p_j)^2 + \nu^2 \left[ \s (q_{j+1},
q_j) - 2 ( \Delta q_j)^2 + \s (q_j, q_{j-1} ) \right]
\label{6.18a}
\end{equation}
Now recall that the normal mode coherent states have the following
special simplifying features: $ \s (q_j, p_j) = 0 $, and the
variances $ (\Delta q_j)^2$, $ (\Delta p_j)^2$ and correlation
functions of the form $ \s (q_{j+1},q_j)$ are independent of $j$.
Also, we find that $C_j = 0 $ (which follows from taking the time derivative
of $\s (q_j, p_j) = 0$). We therefore have
\begin{eqnarray}
\la n(k) \ra &=& \sum_j \exp \left( i k \la q_j\ra - \half k^2 (\Delta q )^2 \right)
\label{6.15a} \\
\la g(k) \ra &=& \sum_j \la p_j \ra
\exp \left( i k \la q_j\ra - \half k^2 (\Delta q)^2 \right)
\label{6.16a} \\
\la \tau (k) \ra &=& \tau_p (k) + \tau_q (k)
\label{6.17a}
\end{eqnarray}
where
\begin{eqnarray}
\tau_p (k) &=&
\sum_j \frac {1} {m} \la p_j \ra^2
\exp \left( i k \la q_j\ra - \half k^2 (\Delta q)^2 \right)
\\
\tau_q (k) &=& \frac {\nu^2} {ik}  \sum_j ( \la q_{j+1} \ra - 2 \la q_j \ra + \la q_{j-1} \ra)
\exp \left( i k \la
q_j\ra - \half k^2 (\Delta q)^2 \right)
\label{6.17b}
\end{eqnarray}
Generally, we do not expect to derive interesting hydrodynamic equations except
in the long wavelength regime. Clearly in this case, this means $ k^{-2} >>
(\Delta q)^2 $. Going to this regime (whose significance will become apparent below),
and reverting back to configuration space, we find
\begin{eqnarray}
\la n(x) \ra &=& \sum_j \delta ( \la q_j \ra - x )
\label{6.19} \\
\la g(x) \ra &=& \sum_j \la p_j \ra \delta ( \la q_j \ra - x )
\label{6.20} \\
\tau_p (x)  &=& \sum_j \frac {1} {m} \la p_j \ra^2
\delta (\la q_j \ra - x )
\label{6.21}
\end{eqnarray}
The quantity $\tau_q (x)$ is more complicated and will be dealt with below.
In these expressions, since we are in the long-wavelength regime, the $\delta$-functions
are to be thought of as smeared over a volume greater than $\Delta q$.

It is now very useful to introduce the velocity field $v(x)$, defined via
\begin{equation}
\la p_j \ra = m v ( \la q_j \ra ) = \int dy \ m v(y) \delta ( \la q_j \ra - y )
\label{6.22}
\end{equation}
which inserted in Eq.(\ref{6.20}) yields
\begin{equation}
\la g(x) \ra = m v(x) \la n (x) \ra
\label{6.23}
\end{equation}
(This is in fact the standard relation between the velocity field and the
momentum density \cite{Hua}).
Inserted also in Eq.(\ref{6.21}) we obtain
\begin{equation}
\tau_p (x)  = m v^2 (x) \la n (x) \ra
\label{6.24}
\end{equation}
It is for deriving these last two equations that the long wavelength assumption
is necessary. Most importantly, $ \tau_p (x) $ is expressed in terms
of the two fields $ v(x)$, $\la n(x) \ra$, which is crucial for closure of the
equations.

We need now to obtain a similarly simple expression for $\tau_q (x) \ra$.
As it stands, Eq.(\ref{6.17b}) will not lead to a simple expression in
terms of $v(x)$ and $\la n(x) \ra$. To proceed further with this term
we need to make simplifications. We are ultimately interested in deriving
a wave equation for the number density (which one might expect on the basis
of the classical equations of motion for the $q_j$). The key to this is
to consider {\it small} displacements of the $q_j$ about uniformly
distributed initial values, and then to consider the fluctuation in number
density about the constant background.
We therefore write
\begin{equation}
\la q_j \ra = j d + \delta q_j
\end{equation}
where $d$ is a constant representing the spacing between each particle
and $\delta q_j$ is a small displacement. The average
number density for $k^{-2}
\gg (\Delta q)^2 $ and to linear order in $\delta q_j$
is then
\begin{eqnarray}
\la n(k) \ra &=& \sum_j e^{ i k \la q_j \ra}
\\
&=& n_0 (k) + n_1(k)  + \cdots
\end{eqnarray}
where
\begin{eqnarray}
n_0 (k) &=& \sum_j e^{ i k j d }
\\
n_1 (k) &=& i k \sum \delta q_j \ e^{ i k j d}
\end{eqnarray}
and note that $n_0 (k) $ is constant. Inserted in the expression for
$ \tau_q (k) $, Eq.(\ref{6.17b}), and assuming also
that $ k^{-1} \gg d $, we find
\begin{equation}
\tau_q (k) = d^2 \nu^2 \ n_1(k)
\end{equation}
This now means that in configuration space we have
\begin{equation}
\la \tau (x) \ra = m v^2 (x) \la n (x) \ra + d^2 \nu^2 \ n_1(x)
\end{equation}

Inserting all of these results in the local conservation equations (\ref{6.1}),(\ref{6.2}),
we obtain a closed set of equations for $ v(x) $ and $\la n(x) \ra = n_0 (x) + n_1(x) $
where
\begin{equation}
n_0 (x) = \sum_j \delta ( x -  j d )
\end{equation}
is a fixed background field (again interpreted as coarse-grained over a length of
order $\Delta q $).
Explicitly, we have
\begin{eqnarray}
\frac {\partial} {\partial t}
\la  n  \ra &=& -   \frac {\partial} {\partial x} ( \la n \ra v )
\label{6.25} \\
\frac {\partial} {\partial t} ( \la n \ra v ) &=& -  \frac
{\partial} {\partial x} \left( \la n \ra v^2 + \frac {d^2  \nu^2}
{m}  n_1 \right) \label{6.26}
\end{eqnarray}
The interesting special case is that in which the velocity field is small,
in which case we may neglect the $v^2$ term in Eq.(\ref{6.26}). Then
combining the two equations yields the wave equation for $ n_1 (x)  $,
\begin{equation}
\frac {\partial^2  n_1 } {\partial t^2} = c^2
\frac {\partial^2  n_1 } {\partial x^2}
\end{equation}
where $c^2 = d^2 \nu^2 / m $.

\subsection{Local Equilibrium and Hydrodynamic Equations for the $ K \ne 0 $ Case}

We now consider the case of the infinite chain with $K \ne 0$. In this case, we expect
that the local density eigenstates will settle down to a local equilibrium state
after a period of time. We will justify this important step below, but first we consider the
consequences of a local equilibrium state, the standard assumption in derivations of
hydrodynamics. This state
is characterized by the one-particle distribution function (Wigner function)
\begin{equation}
w_j (p_j,q_j) = f(q_j) \exp \left( - \frac { (p_j- m v(q_j))^2 } {2 m k T(q_j)} \right)
\label{6.11}
\end{equation}
where $f$, $v$ and $T$ are slowly varying functions of space and time. (This is the one-particle
distribution function for particle $j$ -- it is labelled by $j$ since the particles are distinguishable).
If we now average
the system Eqs.(\ref{6.1})--(\ref{6.3}), together with the currents $\tau (x)$, $j(x)$
in the local equilibrium state, we obtain a closed
system, since we get three equations for three unknowns. In the case of negligible interactions
and $b_j = 0$, we find
\begin{eqnarray}
\la n(x) \ra &=& N f(x)
\label{6.12} \\
\la  g (x) \ra &=& m v(x) N f(x)
\label{6.13} \\
\la h(x) \ra &=& \left(  \half m v^2 + \half k T + \half K x^2 \right) N f (x)
\label{6.14} \\
\la \tau (x) \ra &=&  \left( mv^2 + k T \right) N f(x)
\label{6.15} \\
\la j(x) \ra &=& \left( \frac {3} {2} v k T  + \half m v^3 \right) N f(x) + \frac {K} {2m} x^2 \la g(x) \ra
\label{6.16}
\end{eqnarray}
Inserted in Eqs.(\ref{6.1})--(\ref{6.3}), the above relations give
a closed set of equations for the three variables $f$, $v$ and
$T$. After some rearrangement, these equations are
\begin{eqnarray}
\frac {\partial f} {\partial t} + v \frac {\partial f} {\partial x}
&=& - f \frac {\partial v} {\partial x}
\\
\frac {\partial v} {\partial t} + v \frac {\partial v} {\partial x}
&=& - \frac {1} {m } \frac {\partial \theta } {\partial x}
-\frac {\theta} {m f} \frac {\partial f } { \partial x}
 - \frac {K x } {m}
\\
\frac {\partial \theta } {\partial t} + v \frac {\partial \theta } {\partial x}
&=& - 2 \theta \frac {\partial v} {\partial x}
\end{eqnarray}
where $\theta = k T $. These are the equations for a one-dimensional fluid moving
in a harmonic potential \cite{Hua}. Note that non-trivial equations are obtained even
though we have neglected the interaction terms in deriving them. The role
of interactions is to ensure the approach to local equilibrium,
as we discuss below.

In these expressions, the definition of the velocity field is
equivalent to Eq.(\ref{6.22}) and, similarly, the definition of
the temperature fields is essentially equivalent to,
\begin{equation}
\sum_j \frac {1} {2m} ( \Delta p_j)^2  \ \delta (q_j - x) = \half
k T(x) n (x) \label{6.18}
\end{equation}
(recalling that we are working at long wavelengths, so the
$\delta$-function is coarse-grained over a scale of order
$k^{-1})$. Hence temperature arises not from an environment, but
from the momentum fluctuations averaged over a coarse-graining
volume.

\subsection{The Approach to Local Equilibrium}

Now the key point is that the states we are actually interested in are the approximate eigenstates
of the local densities, evolved in time, or more precisely, the one-particle distribution
function $w_1$ constructed from those states. (Since $w_1$ is the quantity that will enter in
the computation of any averages of sums of one-particle quantities, such as the local densities).
This is not necessarily the same as the local equilibrium distribution,
Eq.(\ref{6.11}), although they are clearly very similar. The averages of the local densities
in the approximate eigenstates will therefore obey the hydrodynamic equations derived above
as long as we can show that the one-particle distribution function of these states $w_1$
approaches the local equilibrium form Eq.(\ref{6.11}) after some time.

The local equilibrium form Eq.(\ref{6.11}) has $\s(q_j,p_j) = 0$
and all the other averages $ \la p_j \ra $, $ \la q_j \ra $, $(\Delta q_j)^2$ and $ (\Delta p_j)^2$
vary slowly in time and space ({\it i.e.} slowly with $j$). (Clearly $\s (q_j, p_j)$ has to be
zero or small for local equilibrium since it is the time derivative of $(\Delta q_j)^2$ and
$ (\Delta p_j)^2$). Compare this with the approximate eigenstates
for the case $K \ne  0$ and the infinite chain. They are Gaussians, so
their one-particle distribution function $w_1$ is entirely determined by
$ \la p_j \ra $, $ \la q_j \ra $, $(\Delta q_j)^2$, $ (\Delta p_j)^2$ and $\s(q_j, p_j)$.
From Eqs.(\ref{3.25})--(\ref{3.27}), we see that $\s (q_j, p_j) $ grows initially
from zero, but then becomes small at late times, whilst
$ (\Delta q_j)^2$ and $ (\Delta p_j)^2 $ approach a constant plus a slowly decaying
factor. Moreover, $ (\Delta q_j)^2$ and $ (\Delta p_j)^2 $ depend on $j$ only through
their initial values, which as discussed (see Eq.(\ref{3.40}) for example), vary slowly.
In addition, the centres $ \la q_j (t) \ra $
and $ \la p_j (t) \ra $ go into a phase of slow time dependence and limited dependence
on $j$ for times sufficiently long for the Bessel functions to go into their
oscillatory phases. These asymptotic forms are approached on a timescale
$ ( \gamma \Omega )^{-1}$. Therefore, in this case, the local equilibrium form
is indeed achieved at late times, and the average values of the local densities
obey hydrodynamic equations.

The final picture we have is as follows. We can imagine an initial
state for the system which contains superpositions of
macroscopically very distinct states. Decoherence of histories
indicates that these states may be treated separately and we thus
obtain a set of trajectories which may be regarded as exclusive
alternatives each occurring with some probability. Those
probabilities are peaked about the average values of the local
densities. We have argued that each local density eigenstate may
then tend to local equilibrium, and a set of hydrodynamic
equations for the average values of the local densities then follow.
We thus obtain a statistical ensemble of trajectories, each of
which obeys hydrodynamic equations. These equations could be very
different from one trajectory to the next, having, for example,
significantly different values of temperature. (In the most
general case they could even be in different phases, for example
one a gas, one a liquid.)

Decoherence requires the coarse-graining scale
$k^{-1}$ to be much greater than the correlation
length of the local density eigenstates, and the derivation of the
hydrodynamic equations requires $k^{-2} \gg ( \Delta q)^2$. In
brief, the emergence of the classical domain occurs on
lengthscales much greater than any of the scales set by the
microscopic dynamics.

\section{Summary and Discussion}

We have shown that for the linear oscillator chain the local
densities define a decoherent set of histories of sufficiently
coarse-grained scales. The key idea is to split the initial state
into local density eigenstates and show that they are preserved in
form under time evolution. The subsequent probabilities for
histories are peaked about the average values of the local
densities, and the equations of motion for them form a closed set
of hydrodynamic form on sufficiently large scales, provided, in
general, that sufficient time has elapsed for the local density
eigenstates to settle down to local equilibrium.

It is perhaps of interest to comment on the timescales involved.
Decoherence through interaction with an environment involves a
timescale, which is typically exceptionally short. Here, however,
there is no timescale associated with decoherence by approximate
conservation. The eigenstates of the local densities remain
approximate eigenstates for all time. There is, however, a
timescale involved in obtaining the hydrodynamic equations,
namely, the time required for a local density eigenstate to
approach local equilibrium. In this model, this timescale is of
order $ ( \gamma \Omega )^{-1}$ (for the infinite chain in the $K
\ne 0 $ case).

Another interesting general issue is the question of the relative
roles of the decoherence through approximate
conservation considered here versus decoherence through
interaction with an environment. The point is that it is a
question of lengthscales. We have demonstrated decoherence of the
local densities starting with exact conservation at the largest
lengthscales and then moving inwards. In this way we were able to
prove decoherence without using an environment, for certain sets
of histories at very coarse-grained scales whose probabilities are
peaked about classical paths. However, in general we would like to be
able to assign probabilities to non-classical trajectories. For
example, what is the probability that a system will follow an
approximately classical trajectory at a series of times, but then
at one particular time undergoes a very large fluctuation away
from the classical trajectory? The approach adopted here indicates
that the probability for this is approximately zero, to the level
of approximation used. Yet this is a valid question that we
can test experimentally. It is at this stage that an
environment becomes necessary to obtain decoherence, and indeed it
is frequently seen in particular models that when there is
decoherence of histories due to an environment, decoherence is
obtained for a very wide variety of histories, not just histories
close to classical. It is essentially a question of information.
Decoherence of histories means that information about the
histories of the system is stored somewhere \cite{GH2,Hal6}.
Classical histories need considerably less information to specify
than non-classical ones, and indeed specification of the three
local densities at any time is sufficient to specify their entire
classical histories. This is not enough for non-classical
histories, so an environment is required to store the information.

Note also that ``environment'' need not necessarily refer to
an external environment. It could also include the internal coordinates
not fixed by the coarse-graining. These did not play a role in the
case considered here, but would become important at finer-grained
levels, producing fluctuations about the evolution described by the
hydrodynamic equations, hand in hand with decoherence. This has been
considered in Refs.\cite{BrHa,CaH1}

Given the need for an environment at finer-grained scales, it is
then of interest to ask whether the local densities continue to
have an important role for many-body systems when an environment
becomes necessary for decoherence. Gratifyingly, the answer is
that the local densities, and particularly the number density,
remain the naturally preferred
variables for a many-body system coupled to an external environment, as was
recently shown \cite{DoHa}. It is normally claimed that position
is the preferred variable in environmentally-induced decoherence,
but this is for a single particle coupled to an environment and is
in any case an approximation. For a many-body system coupled to a
scattering environment, with both described by
many-body field theory, it was shown in Ref.\cite{DoHa} that
number density is the naturally decohering variable (with momentum
density, as its time derivative, also entering in a natural way).
Hence there is a smooth match
with environmentally-induced decoherence models as we go to finer
scales.

It would be of interest to generalize to an oscillator
chain with non-linear interactions. This is because in the linear
chain, the energy in each mode is conserved, so there is no
possibility of exchanging energy between modes, and the approach
to local equilibrium is rather artificial.

It would also be of particular interest to look at a gas. Many-body field
theory may be the appropriate medium in which to investigate this,
following the lead of Ref.\cite{DoHa}. The decoherent histories analysis
might confer some
interesting advantages over conventional treatments.
For example, one-particle dynamics of a
gas is described by a Boltzmann equation. One of the assumptions
involved in the derivation of the Boltzmann equation is that the
initial state of the system contains no correlations, which is
clearly very restrictive \cite{Hua}.
However, in the general approach used here
it is natural to break up any
arbitrary initial state into a superposition of local density
eigenstates, and that these may then be treated separately because
of decoherence. The local density eigenstates
typically have small or zero correlations. Hence, decoherence
gives some justification for one of the rather restrictitve
assumptions of the Boltzmann equation.

We have not estimated the degree of decoherence in the models
considered here, although it could be estimated by looking more
closely at the approximations involved in going from exact to
approximate decoherence, described in Section 2. However, there
may be a more rigorous (but more difficult) way of proving the
results of this paper, which would allow the degree of decoherence
to be estimated. This would be to prove a theorem similar to that
proved by Omn\`es for phase space projectors \cite{Omn}. For a
system of $N$ particles with phase space coordinates ${\bf z} =
({\bf p}, {\bf q})$, Omn\`es considered (approximate) projection
operators onto a region $\Gamma$ ($\gg \hbar$) of phase space,
defined by
\begin{equation}
P_{\Gamma} = \int_{\Gamma} d^N z \ | {\bf z} \ra \la {\bf z} |
\end{equation}
where the states $ | {\bf z} \ra $ are some form of phase space
localized states, such as coherent states. He showed that under
certain reasonable conditions, the form of this projector is
approximately preserved under unitary evolution, that is,
\begin{equation}
e^{ \frac {i}{\hbar} H t } P_{\Gamma} e^{ - \frac {i} {\hbar} H t } \approx P_{\Gamma_t}
\end{equation}
where $\Gamma_t $ is the original phase space region evolved along
the classical phase space trajectories. It is easy to see that
this ensures approximate decoherence of coarse-grained phase space
histories and that the probabilities are peaked about classical
phase space paths. The result is therefore very similar in spirit
to the present paper. It seems very plausible that a similar
result may be proved here for projections onto local densities.
That is, we would like to construct a set of projectors onto the
local densities, $P_{ngh}$ say, and then show that they are
approximately mapped into $P_{n_t g_t h_t}$ under unitary
evolution, where $n_t, g_t, h_t$ are related to the initial values
$n,g,h$ by a closed set of evolution equations. Such a result is
not simply obtained by a coarse-graining of the Omn\`es result,
the issue being that $n_t, g_t, h_t $ have to evolve according to
a {\it closed} set of equations, which is not straightforward to
accomplish in general. (The phase space coordinates ${\bf p}$,
${\bf q}$ evolve according to the Hamilton equations, which is
clearly a closed set of equations, but truncations or
coarse-grainings of this set will generally not be closed).
Moreover, the Omn\`es result breaks down when the underlying
classical dynamics is chaotic. The corresponding hydrodynamic
description, however, being coarser-grained, will generally not
be chaotic and does not obviously break down, so this is a potential
advantage of the hydrodynamic approach. These
and related issues will be pursued in a future publication.

\section{Acknowledgements}

I am very grateful to Jim Hartle and Todd Brun for very many discussions
on the topic of this paper over a long period of time and to Todd Brun
for his critical reading of the manuscript. I would also like
to thank Peter Dodd for useful discussions.




\bibliography{apssamp}

\end{document}